\documentclass[Afour,sageapa,times]{sagej}

\usepackage{moreverb,url}
\usepackage[colorlinks,bookmarksopen,bookmarksnumbered,citecolor=red,urlcolor=red]{hyperref}

\usepackage{multirow}
\usepackage{tabularx}
\usepackage{adjustbox}
\usepackage{xcolor,colortbl}
\usepackage{outlines}
\usepackage{todonotes}
\usepackage{subcaption}
\usepackage{amsmath}
\usepackage{lineno}

\usepackage{subfiles} 

\newcommand\BibTeX{{\rmfamily B\kern-.05em \textsc{i\kern-.025em b}\kern-.08em
T\kern-.1667em\lower.7ex\hbox{E}\kern-.125emX}}

\def\volumeyear{2024}

\def\includedSubjects{65}
\def\male{18}
\def\female{45}
\def\nonbinary{2}

\def\games{41}
\def\amountGames{1 and 5 hours per week}
\def\avgAge{19.7}
\def\stdAge{1.5}

\def\includedGrasps{556}
\def\totalGrasps{2061}

\def\ratedAccuracy{85}
\def\ratedBehavior{15}

\begin{document}

\runninghead{Dekarske et al.}

\title{Anytime Trust Rating Dynamics in a Human-Robot Interaction Task}

\author{Jason Dekarske\affilnum{1}, Gregory Bales\affilnum{1}, Zhaodan Kong\affilnum{1}, and Sanjay Joshi\affilnum{1}}
\affiliation{\affilnum{1}University of California, Davis, Davis, CA, USA}

\email{jdekarske@ucdavis.edu}

\begin{abstract}
\subsubsection{Objective}
We model factors contributing to rating timing for a single-dimensional, any-time trust in robotics measure.

\subsubsection{Background}
Many studies view trust as a slow-changing value after subjects complete a trial or at regular intervals. Trust is a multifaceted concept that can be measured simultaneously with a human-robot interaction.

\subsubsection{Method}
\includedSubjects{} subjects commanded a remote robot arm in a simulated space station. The robot picked and placed stowage commanded by the subject, but the robot's performance varied from trial to trial. Subjects rated their trust on a non-obtrusive trust slider at any time throughout the experiment.

\subsubsection{Results}
A Cox Proportional Hazards Model described the time it took subjects to rate their trust in the robot. A retrospective survey indicated that subjects based their trust on the robot's performance or outcome of the task. Strong covariates representing the task's state reflected this in the model.

\subsubsection{Conclusion}
Trust and robot task performance contributed little to the timing of the trust rating. The subjects' exit survey responses aligned with the assumption that the robot's task progress was the main reason for the timing of their trust rating.

\subsubsection{Application}
Measuring trust in a human-robot interaction task should take as little attention away from the task as possible. This trust rating technique lays the groundwork for single-dimensional trust queries that probe estimated human action.

\subsection{Précis}
When human subjects are given the discretion to rate their trust in a robot counterpart in a single dimension, the timing at which they rate their trust reveals the factor that influences their trust. In this simulated task, a robot's reliability affects its performance, with which subjects change their trust.
\end{abstract}

\keywords{Autonomous agents, Human-automation interaction, Trust in automation, Decision making, Computer interface} 

\maketitle


\section{Introduction}\label{introduction}

Autonomous systems are becoming increasingly capable in daily life and high-risk situations. Each new ``self-driving" car model adds features that remove the need for the operator's full attention. Aspirational future space habitats include integrated autonomous robotic systems to perform work while humans are away \cite{smith_isaac_2021}. The capabilities of these systems introduce a new paradigm for how humans interact with them \cite{chiou_trusting_2023}, including human affective responses like trust.

Measuring and modeling human trust has taken many forms since Mayer's organizational interpretation \cite{mayer_integrative_1995} or Lee's allocation strategies \cite{lee_trust_1992}. Nevertheless, what has stayed constant is the understanding that many factors contribute to how people perceive and interact with autonomous or automated technology. Trust, therefore, is a valuable tool to encapsulate some notion of how human-action relates to robot-action.


Subjects in this study command a remote robot manipulator to sort stowage onboard a simulated International Space Station. The robot's task performance changes from trial to trial according to its ``algorithm" identity. The interface presents a one-dimensional trust slider alongside a rendered video of the real-time simulation. Subjects are free to rate their trust throughout the task at any time.

Given the subjects' freedom to rate their trust at any time, events that coincide with ratings in time are likely to provide the context for why subjects rated their trust. We conducted a survival analysis demonstrating the probability that a subject rates their trust in time relative to a single robotic pick and place action.

\section{Background}\label{background}
Integrating autonomous systems into human work environments has the potential to increase productivity, but with the risk that the system potentially fails. Reliance on these systems is well studied and connects notions of human trust with factors from the system, humans, and the environment. Modulating the level of automation \cite{parasuraman_model_2000} may be a helpful tool to balance performance and risk. Given the affective connection between automation reliance and trust \cite{parasuraman_humans_1997}, capturing a measure of trust throughout interaction makes sense. Trust surveys are a generally accepted measurement mode, but depending on the form and frequency, they can be obtrusive \cite{kintz_estimation_2023}.

Many factors impact one's trust in an autonomous agent, including dispositional factors or situational factors \cite{hoff_trust_2015}. The trust model one uses should be aligned to the timescale for which the conditions change \cite{rodriguez_rodriguez_review_2023}. The collaborative nature of short-time scale interaction may introduce changes in human reliance behavior that do not directly involve the change in a human's trust attitude \cite{lee_trust_2004} and thus requires a separate measure from reliance itself.

The autonomous system in this experiment is a six-degree-of-freedom manipulator, which resembles a human arm with an analogous shoulder, elbow, wrist, and fingers. Since trust is affected by the behavior and appearance of a collaborating autonomous agent \cite{cohen_anthropomorphism_2023}, it is essential to recognize that the identity of the robot, in this case, also plays a role \cite{williams_deconstructed_2021}. We have shown in previous work \cite{dekarske_human_2021} that using identities compartmentalized by ``algorithm" and color was sufficient for subjects to maintain consistent trust evaluations.


%

\subsection{Human decision-making processes}

Studying the human mind from the top down provides a modeling framework that can often describe their behavior. A computational model uses an information processing approach to explain the inputs and outputs of a given system \cite{bermudez_cognitive_2020}. In a human-autonomy teaming task, a human may use information acquired from their senses to make decisions on which actions to take to complete the task. Tourangeau describes that attitudes, such as trust, are ``structures in long-term memory," an information storage module ready to be retrieved and applied in a relevant decision-making context \cite{tourangeau_cognitive_1988}. Reliance decisions and change in trust may not coincide due to the current state of the task or environment. This study aims to investigate trust changes in time relative to current robot actions.



Earlier decision-making models used a utilitarian representation of decision-making where rational agents aimed to maximize some value \cite{sun_computational_2023}. Then, computational models used combinations of strategies and heuristics \cite{kirsch_unifying_2019} to explain behavior. Eventually, information processing modules coalesced to form cognitive architectures like ACT-R \cite{anderson_act_1996}.

Specific decisions can be represented probabilistically using sequential sampling models that integrate information over time as a random walk toward some possible decision threshold. Drift Diffusion Models are a form of sequential sampling that have been useful for investigating decisions involving perception, where one is often required to discriminate between different stimuli \cite{ratcliff_individual_2015}. Huang has demonstrated a drift diffusion model in a quadrotor control task where subjects were asked to move robots to a target position and avoid obstacles \cite{huang_human_2020}. The model was parameterized with robot state error, compared to the goal, as the drift rate, which estimated human decision timing.

Decision Field Theory uses diffusion but models decision preferences as separately sampled stochastic signals that eventually cross a boundary at which time a decision is made. This has been used to model reliance on automation in \cite{gao_extending_2006}. In a pasteurization control task, they found that reliance depends on an interaction between trust and self-confidence, given that a fault occurs in the system.


\subsection{Types of trust measurement}

Although studies have shown a close tie between trust (as a behavior) and automation reliance, trust surveys still prove valuable as a quick attitude assessment. This is especially true when connecting trust to a specific behavior or trait of an autonomous system in a particular situation. Although periodic surveys capture trust well, they cannot identify coincident events that may cause a change in trust.

The semi-automatic pasteurization plant is the canonical environment that measures trust after a fault is introduced in the system \cite{lee_trust_1992}. An autoregressive model describes the response to a series of questions post-trial, the final asking, ``Overall, how much do you trust the system?". Razin and Feigh report that 12.6\% of trust surveys use a single item trust evaluation \cite{razin_converging_2023} and that Jian et al. \cite{jian_foundations_2000} (a multi-question survey) have the most cited survey in use for experiments in the automation field. 

Single-item trust surveys cannot capture the richness and complexity of factors that contribute to the construct \cite{hoff_trust_2015}. However, given the small amount of time needed to complete one, they may be valuable for ``momentary assessments" \cite{bagnara_theoretical_2019}. Proponents of multi-dimensional trust surveys may claim that the lack of detail may not be able to explain the contributing factors to a subject's trust in the system. However, given the interaction context with the subject's freedom to rate, their trust may reveal that these are connected in time.

Reliance in automated systems is a measurable trust factor \cite{rodriguez_rodriguez_review_2023}. Some studies force subjects to chose to rely on an autonomous system at regular intervals - between trials \cite{bhat_clustering_2022} - while others allow reliance at any time - like a driving takeover task \cite{liu_clustering_2021}. On the other hand, a majority of experiments survey trust at regular intervals. Desai et al. require subjects to rate changes in trust during the task at regular intervals \cite{desai_impact_2013}, as well as the OPTIMo study \cite{xu_optimo_2015}. 

Connecting trust to individual features of an interaction requires a subject to decide to rate their trust. Subjects rated their trust at their discretion in \cite{guo_modeling_2021} after a training period. If a robot behaves erratically and then succeeds in the task, an anytime trust survey localizes the trust factor immediately if a subject rates their trust after the robot's behavior. In this case, a short, single-item survey is preferable to limit the disruption and workload of an interactive task.

\section{Experiment}\label{experiment}
\subsection{Experimental task}
Measuring trust in an autonomous robot requires the subject to have some sense of involvement and interaction in the task and robot. The supervisory task we have chosen for this experiment blends risk elements - a spaceflight context - with enough interaction to keep the subject engaged while allowing the robot to contribute to task performance independently.

\begin{figure*}[htb]
    \centering
    \includegraphics[width=\textwidth]{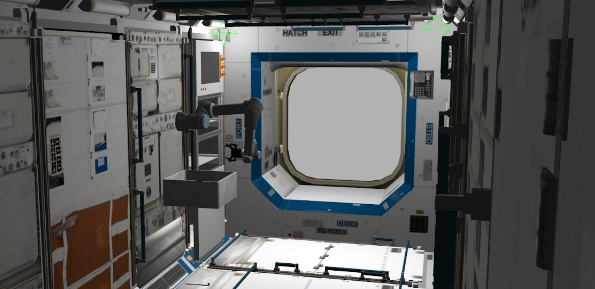}
    \caption[Gazebo Environment]{A rendering of the Gazebo environment where the experiment took place. The subject interacted with the robot through a simulated camera and remote user interface. The simulated UR5e robot arm picked stowage blocks from a container and placed them on the rack in a specified configuration.} 
    \label{fig:simulation}
\end{figure*}

The task simulation occurred in a Gazebo simulation of the International Space Station (ISS) as shown in Figure \ref{fig:simulation} using ROS, a common robotics programming framework. This simulation portrayed a realistic representation of how a physical robot would perform. The robot used a PRM* trajectory planner \cite{kavraki_probabilistic_1996} through a series of consistent gripper waypoints. That way, subjects would see how a robot would move with a true trajectory generator. The anthropomorphic effects of a robot arm could help subjects anticipate the outcome.

\begin{figure*}[ht]
    \centering
    \includegraphics[width=\textwidth]{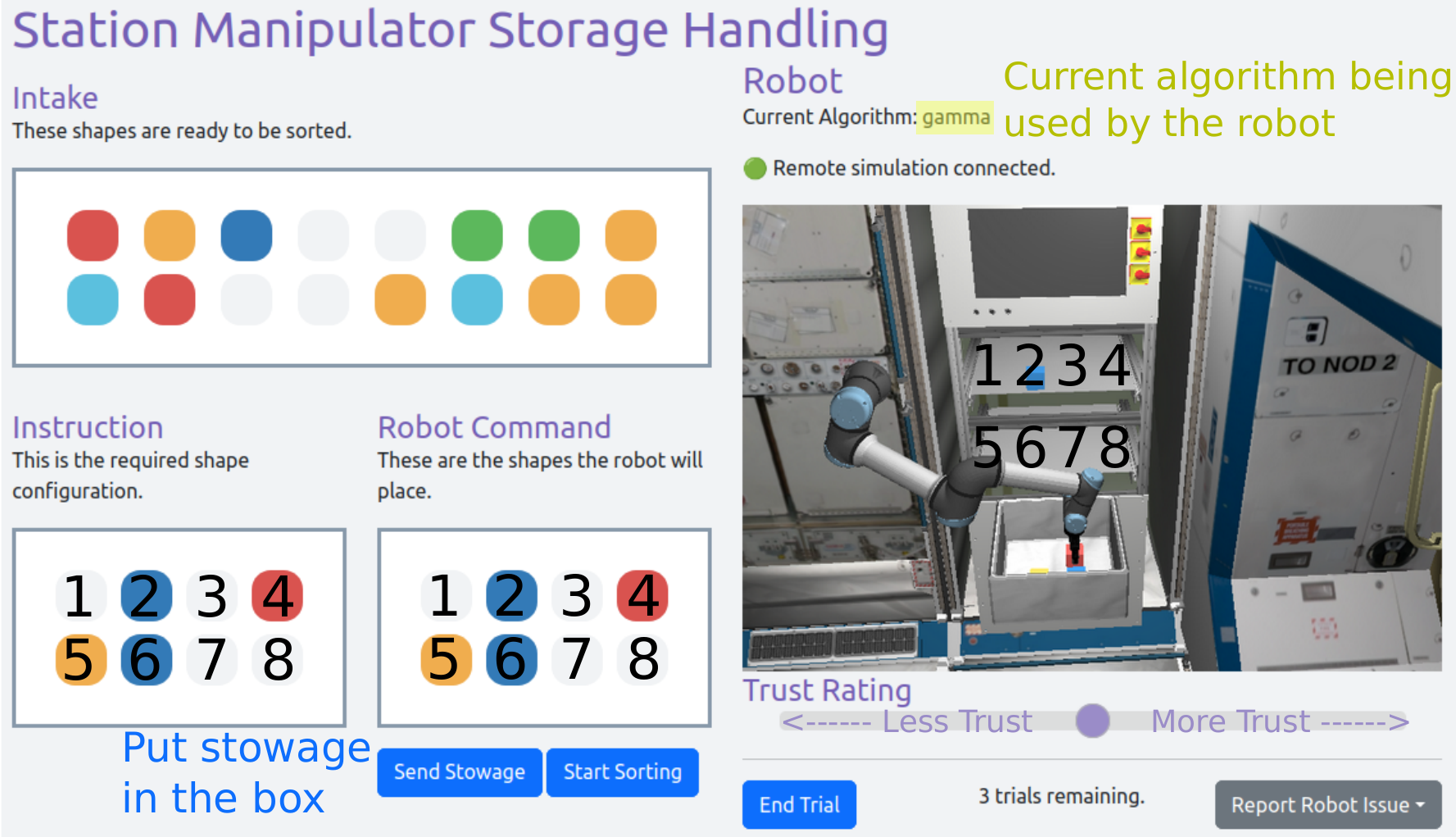}
    \caption[Subject Instructions]{A visual of what subjects see during training. The colored squares correspond to stowage spawned in the box below the rack. Each position shown in \textbf{Instruction} and \textbf{Robot Command}, indicated by number, corresponds to a position shown on the rack. Subjects choose stowage from the \textbf{Intake}, then choose a position in \textbf{Robot Command} to indicate where they want the stowage placed on the rack. The algorithm (Gamma or Echo) is indicated above the robot simulation live feed. The trust rating slider is accessible below the live feed. The markup on this image appeared only for instruction.} 
    \label{fig:instructions}
\end{figure*}

The human-robot team aimed to move proper stowage pieces, represented by colored cubes, to their proper position on a rack. We chose colored cubes to reduce ambiguity about whether the robot successfully completed the task. The subject provided supervisory commands to the robot through their remote user interface. The instructions for the subject overlaid on the remote user interface are shown in Figure \ref{fig:instructions}. Along with this graphic was a description of the goals of the experiment, including an emphasis on ``improving the way people work with smart technology". The experiment was designed to take between 30 and 60 minutes. Subjects were instructed to complete the task with a reliable internet connection and a quiet environment.



To measure the trust the subject had in the robot, we used the following prompt in the instructions:

\begin{quote}
    Throughout the process, please adjust your trust in the system using the purple slider beneath the camera feed where the left indicates less trust and the right indicates more trust.
\end{quote}

The trust slider always retained its position from when the subject released it. Although trust is a major component in this study, we wanted to ensure that the measure of trust would not bias subjects to rate their trust in any specific feature of the robot. Lastly, subjects were provided a disclaimer about the reality of live robotics simulations and that it was possible for the stowage cubes or the robot to display strange behavior.

The robot's performance was modulated as an experimental condition. There are two ``algorithms" used by the robot for stowage placement: \textit{Gamma} and \textit{Echo}. These embodiments separated the task performance of the robot into two distinct categories. Gamma performs correct placements every time, while Echo places stowage in the wrong position 50\% of the time. The current algorithm was shown to the subject on the interface.

\subsection{Experimental design}
\begin{figure}[htb]
    \centering
    \includegraphics[width=\columnwidth]{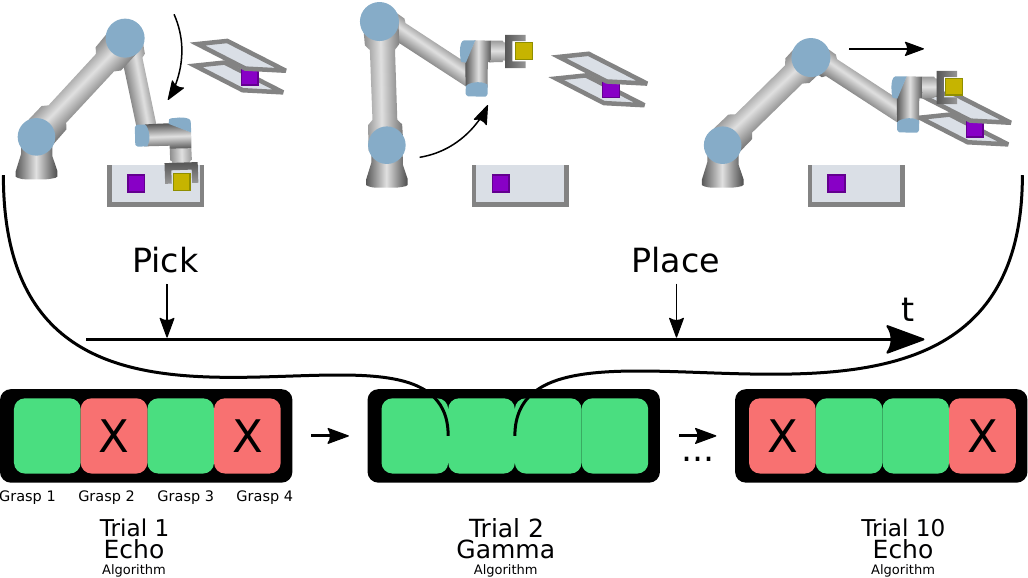}
    \caption[Experiment Block]{Each included subject completes ten trials of 4 grasps each. An algorithm, Echo or Gamma, is assigned randomly for each trial. A grasp is defined as the moment the gripper closes on a piece of stowage to the next time it closes on stowage. This way, trust ratings are associated with the prior grasp. Subjects may adjust their trust rating at any time.}
    \label{fig:design}
\end{figure}

The experiment was split into ten trials of four grasps each. The robot algorithm was randomly assigned each trial to control for possible ordering effects. Figure \ref{fig:design} shows a possible subset of trials for a single experimental subject. Although the robot makes mistakes, the subject is likely to assign blame to the ``algorithm" identity. This way, they may track the robot's reliability throughout the experiment. 

Our hypotheses assume that subjects' trust in the robot is tied to the robot's behavior and performance. To validate this, we asked open questions after the experiment about what influenced them to rate their trust. The trust slider was intentionally optional to not bias the subject into rating their trust according to our hypothesis, but checked upon the completion of the experiment with the following open-ended questions:

\begin{itemize}
    \item What caused you to change your trust in the robot?
    \item Why do you think the robot failed to complete the task at times?
    \item Was any part of the experiment difficult or hard to understand?
    \item Were there any technical problems with the experiment? 
\end{itemize}

The University of California’s Davis Institutional Review Board approved this research, and it complies with the American Psychological Association Code of Ethics. Participants provided consent and were allowed to exit the experiment without penalty.

\section{Trust Model}\label{model}
\subsection{Trust in a decision-making paradigm}
In this experiment, subjects could rate their trust at any time. Subjects' latent trust in the robot changes continuously over time depending on many factors. However, their conscious introspection of their trust state happens when they are prompted or face a situation that defies their expectations. In the case of the interaction shown here, the robot's task performance is the central stimulus contributing to changing trust. Although many factors may change someone's trust in a robot, performance is the most salient over the timescale of this experiment.

Subjects are instructed to ``adjust their trust throughout the process" of the experimental task. The experimental instructions were biased toward completing the sorting task, with trust reporting left as a secondary, optional task. The purpose of this design was not to bias the interpretation of trust to rely on any single factor. Additionally, without forcing trust ratings at regular intervals, the subjects would not expect some stimulus to appear that they should use to rate their trust. The experiment included these details to demonstrate that lay people can introspect their trust state in a robot collaborator without definitions of trust.



As subjects observe the robot progressing through the task, they gain evidence about its behavior, capabilities, performance, and more. Eventually, when they have decided they have enough information, they may realize that their trust has changed. From the Sequential Sampling Model's point of view, they have accrued enough information to meet a threshold and make a decision \cite{sun_computational_2023}. For example, the drift diffusion model (DDM) is used in a task of a \textit{two-alternative forced choice} task. Over time, evidence is accrued as a random walk with a drift rate. The corresponding response is chosen when the signal reaches a specific positive or negative threshold. This model may be more suitable for trust decisions that are binary or have forced decisions.

\subsection{Trust rating in a survival model} \label{sec:survival_model}
When subjects decide to report their trust rating in the robot, they have gathered sufficient evidence. Whether that evidence comes from performance, behavior, or some other covariate is what we aim to model. Assuming the subjects' trust is tied to each \textit{grasp} that a robot attempts, we portion each grasp starting with the moment the robot closes its gripper on a piece of stowage and ending after the robot places stowage before grabbing the next stowage. Since subjects are free to change their trust throughout a grasp, we assume that they have acquired the \textit{necessary} evidence to report their trust by the time the last trust rating occurred during a grasp. This moment is the Trust Rating Time (tRT).

Survival Analysis \cite{cox_analysis_1984} models the expected time it takes for an event to occur. Here, we aggregate all the grasps for which subjects decided to report their changed trust. The survival function is given by the following formulation:

\begin{equation}
    S(t)= Pr(T>t)
\end{equation} 

\noindent
Where $t$ is the time since the start of the measurement period when the robot picks the stowage, and then $T$ is the tRT. The survival function is never increasing since waiting to rate trust later means the \textit{last} trust rating could not have already happened. The function describes the probability that the tRT \textit{survives} past the time $t$.

The complement of the Survival function is the cumulative probability distribution of tRTs.

\begin{equation}
    F(t) = Pr(T \leq t) = 1 - S(t)
\end{equation} 

Estimating the tRT density requires the hazard function $\lambda(t)$, which is the instantaneous rate of an event occurring conditional on surviving to that time.

\begin{equation}
    \lambda(t) = \lim_{t \to 0} \frac{Pr(t \leq T < t + dt)}{dt \cdot S(t)} = \frac{F'(t)}{S(t)} = -\frac{S'(t)}{S(t)}
    \label{eqn:hazard}
\end{equation} 


The Cox Proportional Hazards Model describes a hazard function that includes multiplicative covariates with respect to the baseline hazard, $\lambda_0$, shown in its general form in Equation (\ref{eqn:cox}) \cite{cox_analysis_1984}. This model uses constant and time-varying covariates represented by $x$ and $y$ respectively. Their associated constant coefficients are $\beta$ and $\eta$. In vector $x$, the constant covariates used from this experiment are 1) grasp success and 2) trust. $y$ is the time-varying covariate, grasp completion, which is 0 until the stowage is placed, then 1. 

To solve for the coefficients in the hazard function, we use the Equivalent Poisson Model as described in \cite{rodriguez_survival_2010}. We split each grasp over the experiment, $i$, into equally spaced intervals in time, $j$. If a subject rated trust in an interval, take \textit{rating indicator} $d_{i,j}$ ($d$ is traditionally used as a ``death" indicator in survival analysis) as 1, otherwise 0. Next, \textit{exposure} indicates the intervals for which the covariates interact in the hazard function: $e_{i,j}$ is 1 for each interval $j$ which the subject has not yet rated their trust, or the time in the interval that that they rate their trust, otherwise 0. Then, we take $\mu_{i,j}=e_{i,j}\lambda_{i,j}$ as the expected number of ratings. $\mu_{i,j}$ is fit to the observed rating times in $d_{i,j}$. This formulation allows for time-varying covariates, $y_{i,j}$, in Equation (\ref{eqn:cox_interval}).

\begin{align} 
    \lambda(t)&=\lambda_{0}(t)exp(x\beta) \label{eqn:cox} \\
    \lambda_{i,j}&=\lambda_{0}exp(x_{i}\beta + y_{i,j}\eta) \label{eqn:cox_interval}
\end{align}

To build an extremely well-fit model, one could vary $\lambda_0$ with time or interval to represent a more general form or Weibull baseline \cite{rodriguez_survival_2010}. We chose a constant baseline because the time-varying effects are captured by the progress of the robot's grasp. Additionally, a censored model would be more appropriate if trust ratings \textit{could} have been made in the future. However, this was not appropriate due to the time pressure inherent in the sequence of grasps.

We used PyMC \cite{patil_pymc_2010}, a Bayesian modeling Python library, to fit the parameters in this model. After specifying a model symbolically and assigning priors for parameters, PyMC samples from the model and updates a posterior distribution of parameter values using the No-U-Turn Sampler \cite{hoffman_no-u-turn_2011}.


\section{Results}\label{results}
\subsection{Demographics and exclusion}\label{sec:demographics}
Subjects were excluded from the experiment if they did not finish the entire study, had unusually long trials, or had an unstable internet connection characterized by median latency greater than 300ms. There were \includedSubjects{} total subjects included, of which \male{} identified as male, \female{} female, and \nonbinary{} nonbinary. The average age was \avgAge{} years with standard deviation \stdAge{}. Of the \games{} subjects who played video games, the median subject played between \amountGames{}.

The remote study enabled subjects to participate in the experiment from wherever was comfortable for them. However, this introduces a non-negligible delay in the timing of interactions. Most subjects were recruited from the UC Davis campus but may have completed the experiment remotely. For example, baseline latency to San Francisco is $\approx40ms$, measured using speedtest.net. The user interface tracked latency every 10 seconds by calling the \textit{get\_time} ROS service and measuring the duration before receiving a reply. Then, latency was corrected in analysis by applying a rolling median filter to the latency measurement and shifting the logged data by half the measurement to be conservative.




\subsection{Measuring trust}
\begin{figure}[htb]
  \centering
  \includegraphics[width=\linewidth]{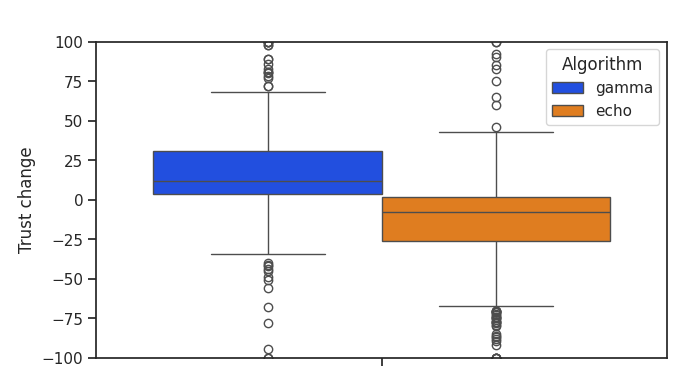}
  \caption[Change in Trust Over Each Grasp]{This comparative box plot shows the change in trust averaged over every grasp in which they changed trust. The capabilities, or algorithms, are shown in different colors where the Gamma algorithm always places stowage correctly, and the Echo algorithm misplaces stowage 50\% of the time. This analysis treats \emph{every} grasp as an independent event.}
  \label{fig:trust_change}
\end{figure}

Subjects changed their trust using the trust slider throughout the experiment. Each included subject completed ten trials, which consisted of 4 grasps each. Figure \ref{fig:trust_change} shows an aggregation of trust change over all the grasps in which subjects rated their trust. The trust change data are summarized in Table \ref{tab:trustchange}.

\begin{table}[htb]
\centering
\begin{tabular}{lrrr}
 & count & mean & std \\
Algorithm &  &  &  \\
echo & 303.0 & -14.7 & 33.1 \\
gamma & 253.0 & 16.3 & 32.1 \\
\end{tabular}

\caption[Average Trust Change]{Average trust change according to algorithm. The slider ranges from 0 to 100. Using a paired t-test, the trust change is significantly different between the Gamma and Echo algorithms ($p\ll0.01$).}
\label{tab:trustchange}
\end{table}

\begin{figure}[htb]
  \centering
  \includegraphics[width=\linewidth]{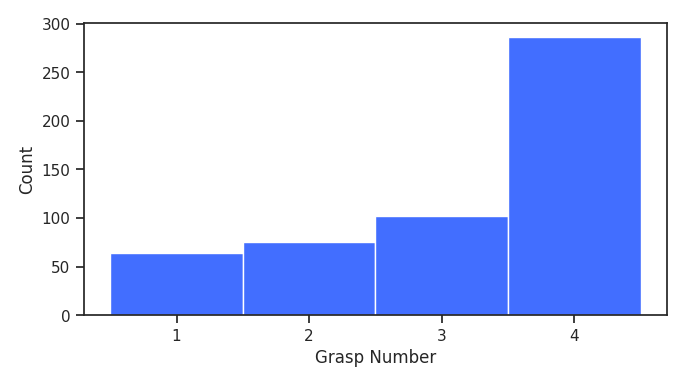}
  \caption[Trust Change Distribution By Grasp]{Distribution of occurrence of subjects' trust ratings by grasp number. Most ratings occurred during the final grasp.}
  \label{fig:median_grasp}
\end{figure}

Although subjects can rate their trust at any time, we found that they typically rated trust towards the end of the trial. This suggests that subjects wanted complete evidence over more grasps before evaluating their trust in the robot. Figure \ref{fig:median_grasp} shows the distribution of the when trust ratings occurred by grasp number.

\subsection{Trust reaction time}
\begin{figure}[htb]
  \centering
  \includegraphics[width=\linewidth]{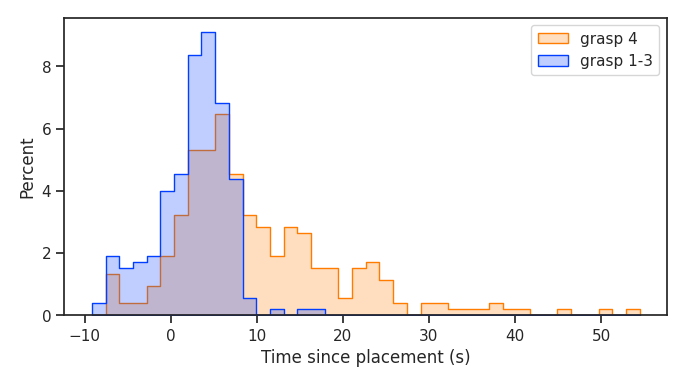}
  \caption[Trust Rating Time Distribution]{Trust rating time distribution separated by whether the grasp was the last one. The \textit{pick} happens near -10s. When the grasp is the last of the trial, subjects have unlimited time to change their trust. For the first three grasps, subjects are pressured to rate their trust before the next grasp near 10s. }
  \label{fig:rt_distribution}
\end{figure}

There is a clear delineation between when a grasp ends and the next one begins. Since each of the four grasps proceeds in sequence, there is a natural time pressure to rate trust, if they choose to, before the start of the next grasp. We see this illustrated in Figure \ref{fig:rt_distribution}. Although the distributions of rating times have similar peaks, there is a noticeable drop-off around 10 seconds for the first three grasps since this is when the next grasp starts. Because of this time pressure, we parameterized the reaction times differently for the first three and the last grasp (four grasps total).

Out of the total \totalGrasps{} grasps witnessed by subjects during the experiment, they rated their trust in \includedGrasps{} of them. Applying the Survival Analysis procedure, we produce the parameters in Figure \ref{tab:parameter_table}. The values have converged as evident from the \textit{r\_hat} statistic, or the Gelman-Rubin Statistic, since the value is nearly close to 1 for each parameter. The parameters between the models for the first and last grasp are comparable. Interestingly, neither trust nor success has an impact compared to $\eta$, the coefficient of the grasp state covariate. However, the \textit{success} coefficient is significant compared to the baseline hazard, the $\lambda_0$ constant, suggesting that the impending \textit{success} of a grasp contributes to a higher probability of trust rating before the grasp has been completed.

\begin{figure}
  \begin{subfigure}[b]{\linewidth}
    \centering
  \begin{tabular}{lrrr}
\toprule
 & mean & sd & r\_hat \\
\midrule
beta[success] & 0.15 & 0.15 & 1.00 \\
beta[trust] & -0.00 & 0.00 & 1.00 \\
eta & 1.97 & 0.16 & 1.00 \\
lambda0 & 0.03 & 0.01 & 1.00 \\
\bottomrule
\end{tabular}

    \caption[Grasp 1-3 Parameters]{Grasps 1-3}
  \end{subfigure}%
  \\
  \begin{subfigure}[b]{\linewidth}
    \centering
  \begin{tabular}{lrrr}
\toprule
 & mean & sd & r\_hat \\
\midrule
beta[success] & 0.07 & 0.14 & 1.00 \\
beta[trust] & -0.00 & 0.00 & 1.00 \\
eta & 2.34 & 0.22 & 1.00 \\
lambda0 & 0.01 & 0.00 & 1.00 \\
\bottomrule
\end{tabular}

    \caption[Final Grasp Parameters]{Final grasp}
  \end{subfigure}
  \caption[Hazard Function Parameters]{Parameter tables for the hazard function described in Equation (\ref{eqn:cox_interval}). \textit{a} and \textit{b} are fit separately because the subjects have no time pressure to register their trust for the final grasp. The $\eta$ coefficient is time-varying, while the $\beta$ coefficients are not. $\lambda0$ is the baseline hazard. All r\_hat values are close to 1, indicating convergence.}
  \label{tab:parameter_table}
\end{figure}

\begin{figure}
  \begin{subfigure}[b]{\linewidth}
    \centering
    \includegraphics[width=\linewidth]{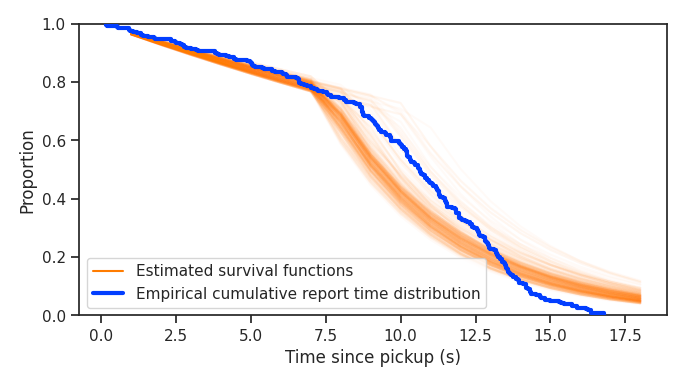}
    \caption[Grasp 1-3 Survival Function]{Grasps 1-3}
  \end{subfigure}%
  \\
  \begin{subfigure}[b]{\linewidth}
    \centering
    \includegraphics[width=\linewidth]{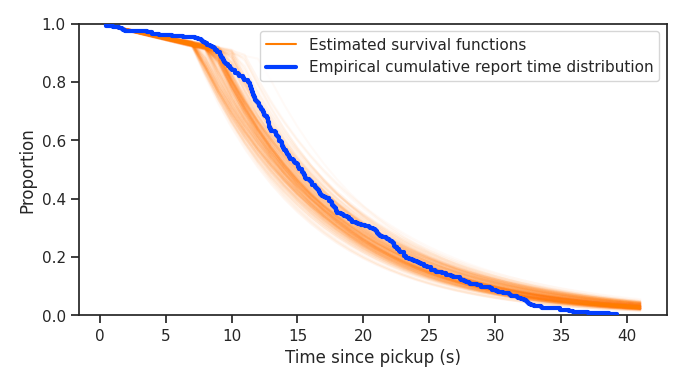}
    \caption[Final Grasp Survival Function]{Final grasp}
  \end{subfigure}
  \caption[Predicted Survival Functions]{Predicted survival functions for the set of covariates from each grasp where subjects rated their trust compared with the empirical survival function. The model fits well for the final grasp. For the first grasps, the model predicted that subjects were likely to rate their trust sooner than what the observations showed. The large change in slope happens when the robot places the stowage on the rack. After this time, the rate at which subjects registered their trust increased dramatically.}
  \label{fig:RT_prediction}
\end{figure}

Given the parameterized hazard function, we can sample the observed covariates and plot a survival function for each one shown in Figure \ref{fig:RT_prediction}. Since most of the grasps had similar placement durations, we can overlay the empirical cumulative distribution function of reaction times for comparison. This function should fall near the center of the predicted traces for a well-fit model.

\section{Discussion}\label{discussion}
\subsection{Deciding to trust}\label{sec:decision}

When a subject decides to rate trust, they have processed enough information about the situation, combined with their previous experience with the robot, to introspect. This experiment aimed to capture the total time from robot action to trust-rating action. The supervisory task required the subject to command robot action. The robot's performance could be readily tracked according to the ``algorithm" it followed. Since this was the most salient trusting attribute, it was likely the basis of the subjects' trust.

Subjects had the freedom to rate their trust at any time during the grasp and were not limited to the conclusion of the placement. Although many waited for this moment, some rated the grasp well before, anticipating the outcome. The timescale with which they waited was on the second's scale, as seen in Figure \ref{fig:rt_distribution}, but some took longer when given the opportunity.


\subsection{Survival analysis and decision making}\label{sec:survival_discussion}

The survival function described in this work illustrates the probability that a subject waits to decide to rate their trust relative to when the robot picks up a piece of stowage. When the robot initially grabs a piece of stowage, the subject must check that it is correct, then observe as the end-effector approaches the right location on the rack. When they are confident that the robot is correct or makes a mistake, they may decide to change their trust.

If they integrate information from their observation, it makes sense that the subject is more sure of the outcome as the robot gets closer to the goal. In other words, the probability that the robot's reliability differs from expectations increases or decreases faster as the outcome approaches. This is akin to the \textit{hazard} accumulating \cite{betancourt_outwit_2022} over time.

\begin{equation}
    \Lambda(t) = \int_{0}^{t} \mathrm{d} t' \, \lambda(t)
\end{equation} 

We can rewrite the final part of equation \ref{eqn:hazard} and solve for the survival function.

\begin{align}
    \lambda(t) &= -\frac{d}{dt}\log(S(t)) \\
    S(t) &= \exp(- \int_{0}^{t} \mathrm{d} t' \, \lambda(t)) = \exp(-\Lambda(t))
\end{align}

In this case, the cumulative hazard, $\Lambda(t)$ is a natural fit for relating stimulus over time to a decision-making event, a trust rating.

Figure \ref{fig:RT_prediction} shows a change in the slope of the predicted survival function around the time the robot places the stowage. Since the proportional hazards model's covariates are multiplicative over the baseline hazard, any differences in parameters in Figure \ref{tab:parameter_table} describe differences in the rate of the probability of trust rating. For both sets of parameters, $\eta$, the time-varying state of the grasp, is much larger than the baseline hazard, indicating the hazard increases dramatically after the grasp completes. The baseline $\lambda_0$ for the earlier grasps is larger than the one for the final grasp, which mainly corresponds to the risk before the placement occurs, which means the probability of rating trust before stowage is higher before the last grasp.

The model proposed here successfully converged to a set of parameters that captured the shape of the empirical cumulative distribution of trust rating times. The time pressure of the first three grasps influenced subject behavior in a way that could not be captured with this model.
The ``time until the next grasp" is information not accessible until after the next grasp starts, making it invalid for this analysis. Allowing subjects to control the start of individual grasps would eliminate time pressure to rate trust and align the behavior seen during the final grasp of each trial.

A single-item trust survey may prove useful for assessing an overall feeling or impression. However, more information is needed if subjects fail to consider the complete set of possible factors that influence their decision \cite{bagnara_theoretical_2019}. The experimental task's most salient feature was the outcome of each pick and place. The proposed model identifies this feature as the most influential and thus sufficient for this analysis. Since task outcome changes throughout the experiment, it makes sense that a subject is most attentive to this factor. However, suppose a researcher was interested in other factors like predictability, familiarity, or developer intention. In that case, a multidimensional survey may be more appropriate \cite{bagnara_theoretical_2019}. Capturing these other factors requires the appropriate covariates from the task.

\subsection{Recommendations for a one dimensional trust slider}\label{sec:recommendations}

An exit survey was administered after the experiment, which consisted of free-response questions designed to learn how subjects interacted with the robot. When asked what caused them ``to change their trust in the robot?" \ratedAccuracy \% of responses indicated that either ``accuracy" or ``stowage placement" affected their assessment; \ratedBehavior \% indicated that robot behavior affected their trust (not mutually exclusive). Although subjects were not provided a trust definition, their explanation aligned with our expectation that trust would be related to the robot's task performance.

Of the responses that were not accuracy or behavior-related, two mentioned the robot not listening to their commands; one tested how changing the slider could affect the robot, one stated, ``It had no mind," and one was unsure what trust meant. A large majority of subjects indicated post hoc that they based their trust on the outcome of the task without prompting, which suggests that the trust slider method captures the volitional trust rating we aimed for.

Our subject pool was limited to undergraduate students at an English-speaking university. If a similar study uses this survey concept, researchers should consider how the colloquial definition of trust changes across language and education status.

Throughout the experiment, the trust slider never reset to the central starting position; it retained the value with which it was last set. After a grasp, trial, or algorithm change, the slider could have been reset but may have prompted the subject to rate their trust at that time rather than when they felt their trust had changed. If a researcher wished to receive more trust ratings at the expense of altering the subject's attention, they may want to reset the slider and prompt the subject to rate their trust.

\subsection{Operationalizing any-time trust rating}

We envision an interactive system that uses trust queries to capture part of humans' complex decision-making process. If a robot, for example, could ask a collaborating human about their trust in itself, it could have an idea of what the human will likely do in the future \cite{dissing_implementing_2020}. This study highlights the possibility of an unobtrusive single-dimensional trust measure that captures reactions to a robot's task performance and behavior. This measure may prove valuable as human-robot relationships transition from a supervisory hierarchy to a flattened decision-making structure \cite{chiou_trusting_2023}.

\section{Conclusion}\label{conclusion}
Subjects use evidence to rate their trust in the robot. Their \textit{decision} requires them to contemplate what aspects of the robot in this context change their trust. There is some probability that subjects rate their trust before the robot places the stowage because they can anticipate the outcome or evaluate robot behavior. We provide the trust survey as a single-dimensional slider, but since subjects can rate trust at their discretion, the factors that influence their trust will likely coincide with the rating. Additionally, the generality of the survey allows subjects to emphasize which factor of trust most affected their trust report.

This model did not take individual demographics or individual behavior into account. After some interaction with the experiment, learned behavior could have changed when the subject decided to rate trust after knowing the robot's motion; the subject could anticipate the outcome sooner. According to the model, trust and task outcome played little role in the actual timing of subjects' trust reports, only that the timing of the task outcome coincided with the trust report. 

The findings from this study can be used to inform future any-time trust rating experiment designs. Most subjects reported task-performance-related reasons for changing their trust, but the supervisory nature of the robotic task likely influenced this; for more collaborative tasks, subjects may report other factors for changing their trust. With the increasing intersection of humans and autonomous systems, quick trust queries like the survey presented in this study could enable more fluid interactions.

\begin{acks}
This material is based upon work supported by NASA (award number 80NSSC19K1052). Any opinions, findings, and conclusions or recommendations expressed in this material are those of the author(s) and do not necessarily reflect the views of the National Aeronautics and Space Administration (NASA).
\end{acks}

\section{Key Points}
\begin{itemize}
    \item A single-dimensional trust self-report slider administered continuously throughout a human-robot interaction experiment found changes in trust in line with the performance of the robotic system.
    \item Survival Analysis proves a valuable tool for evaluating covariates, which identify factors subjects use to decide when to report their trust changed.
    \item In this experiment, the state of the task, when the robotic agent could be assessed for its performance, was the most impactful factor for trust rating.
\end{itemize}




\bibliographystyle{apalike}
\bibliography{picknplace.bib}

\begin{thebibliography}{}

\bibitem[Anderson, 1996]{anderson_act_1996}
Anderson, J.~R. (1996).
\newblock {ACT}: {A} simple theory of complex cognition.
\newblock {\em American Psychologist}, 51(4):355--365.

\bibitem[Bermudez, 2020]{bermudez_cognitive_2020}
Bermudez, J.~L. (2020).
\newblock {\em Cognitive {Science}: {An} {Introduction} to the {Science} of the {Mind}}.
\newblock Cambridge University Press, 3 edition.

\bibitem[Betancourt, 2022]{betancourt_outwit_2022}
Betancourt, M. (2022).
\newblock Outwit, {Outlast}, {Outmodel}.
\newblock Technical report, .

\bibitem[Bhat et~al., 2022]{bhat_clustering_2022}
Bhat, S., Lyons, J.~B., Shi, C., and Yang, X.~J. (2022).
\newblock Clustering {Trust} {Dynamics} in a {Human}-{Robot} {Sequential} {Decision}-{Making} {Task}.
\newblock {\em IEEE Robotics and Automation Letters}, 7(4):8815--8822.

\bibitem[Chiou and Lee, 2023]{chiou_trusting_2023}
Chiou, E.~K. and Lee, J.~D. (2023).
\newblock Trusting {Automation}: {Designing} for {Responsivity} and {Resilience}.
\newblock {\em Human Factors: The Journal of the Human Factors and Ergonomics Society}, 65(1):137--165.

\bibitem[Cohen et~al., 2023]{cohen_anthropomorphism_2023}
Cohen, M.~C., Peel, M.~A., Scalia, M.~J., Willett, M.~M., Chiou, E.~K., Gorman, J.~C., and Cooke, N.~J. (2023).
\newblock Anthropomorphism {Moderates} the {Relationships} of {Dispositional}, {Perceptual}, and {Behavioral} {Trust} in a {Robot} {Teammate}.
\newblock {\em Proceedings of the Human Factors and Ergonomics Society Annual Meeting}, 67(1):529--536.

\bibitem[Cox and Oakes, 1984]{cox_analysis_1984}
Cox, D.~R. and Oakes, D. (1984).
\newblock {\em Analysis of survival data}.
\newblock Monographs on statistics and applied probability. Chapman and Hall, London ; New York.

\bibitem[Dekarske and Joshi, 2021]{dekarske_human_2021}
Dekarske, J. and Joshi, S.~S. (2021).
\newblock Human {Trust} of {Autonomous} {Agent} {Varies} {With} {Strategy} and {Capability} in {Collaborative} {Grid} {Search} {Task}.
\newblock In {\em 2021 {IEEE} 2nd {International} {Conference} on {Human}-{Machine} {Systems} ({ICHMS})}, pages 1--6, Magdeburg, Germany. IEEE.

\bibitem[Desai et~al., 2013]{desai_impact_2013}
Desai, M., Kaniarasu, P., Medvedev, M., Steinfeld, A., and Yanco, H. (2013).
\newblock Impact of robot failures and feedback on real-time trust.
\newblock In {\em 2013 8th {ACM}/{IEEE} {International} {Conference} on {Human}-{Robot} {Interaction} ({HRI})}, pages 251--258, Tokyo, Japan. IEEE.

\bibitem[Dissing and Bolander, 2020]{dissing_implementing_2020}
Dissing, L. and Bolander, T. (2020).
\newblock Implementing {Theory} of {Mind} on a {Robot} {Using} {Dynamic} {Epistemic} {Logic}.
\newblock In {\em Proceedings of the {Twenty}-{Ninth} {International} {Joint} {Conference} on {Artificial} {Intelligence}}, pages 1615--1621, Yokohama, Japan. International Joint Conferences on Artificial Intelligence Organization.

\bibitem[Gao and Lee, 2006]{gao_extending_2006}
Gao, J. and Lee, J. (2006).
\newblock Extending the decision field theory to model operators' reliance on automation in supervisory control situations.
\newblock {\em IEEE Transactions on Systems, Man, and Cybernetics - Part A: Systems and Humans}, 36(5):943--959.

\bibitem[Guo and Yang, 2021]{guo_modeling_2021}
Guo, Y. and Yang, X.~J. (2021).
\newblock Modeling and {Predicting} {Trust} {Dynamics} in {Human}–{Robot} {Teaming}: {A} {Bayesian} {Inference} {Approach}.
\newblock {\em International Journal of Social Robotics}, 13(8):1899--1909.

\bibitem[Hoff and Bashir, 2015]{hoff_trust_2015}
Hoff, K.~A. and Bashir, M. (2015).
\newblock Trust in {Automation}: {Integrating} {Empirical} {Evidence} on {Factors} {That} {Influence} {Trust}.
\newblock {\em Human Factors: The Journal of the Human Factors and Ergonomics Society}, 57(3):407--434.

\bibitem[Hoffman and Gelman, 2011]{hoffman_no-u-turn_2011}
Hoffman, M.~D. and Gelman, A. (2011).
\newblock The {No}-{U}-{Turn} {Sampler}: {Adaptively} {Setting} {Path} {Lengths} in {Hamiltonian} {Monte} {Carlo}.

\bibitem[Huang et~al., 2020]{huang_human_2020}
Huang, J., Wu, W., Zhang, Z., and Chen, Y. (2020).
\newblock A {Human} {Decision}-{Making} {Behavior} {Model} for {Human}-{Robot} {Interaction} in {Multi}-{Robot} {Systems}.
\newblock {\em IEEE Access}, 8:197853--197862.

\bibitem[Jian et~al., 2000]{jian_foundations_2000}
Jian, J.-Y., Bisantz, A.~M., and Drury, C.~G. (2000).
\newblock Foundations for an {Empirically} {Determined} {Scale} of {Trust} in {Automated} {Systems}.
\newblock {\em International Journal of Cognitive Ergonomics}, 4(1):53--71.

\bibitem[Johnson and Busemeyer, 2023]{sun_computational_2023}
Johnson, J.~G. and Busemeyer, J.~R. (2023).
\newblock Computational {Models} of {Decision} {Making}.
\newblock In Sun, R., editor, {\em The {Cambridge} {Handbook} of {Computational} {Cognitive} {Sciences}}, pages 499--526. Cambridge University Press, 2 edition.

\bibitem[Kavraki et~al., 1996]{kavraki_probabilistic_1996}
Kavraki, L., Svestka, P., Latombe, J.-C., and Overmars, M. (1996).
\newblock Probabilistic roadmaps for path planning in high-dimensional configuration spaces.
\newblock {\em IEEE Transactions on Robotics and Automation}, 12(4):566--580.

\bibitem[Kintz et~al., 2023]{kintz_estimation_2023}
Kintz, J.~R., Banerjee, N.~T., Zhang, J.~Y., Anderson, A.~P., and Clark, T.~K. (2023).
\newblock Estimation of {Subjectively} {Reported} {Trust}, {Mental} {Workload}, and {Situation} {Awareness} {Using} {Unobtrusive} {Measures}.
\newblock {\em Human Factors: The Journal of the Human Factors and Ergonomics Society}, 65(6):1142--1160.

\bibitem[Kirsch, 2019]{kirsch_unifying_2019}
Kirsch, A. (2019).
\newblock A unifying computational model of decision making.
\newblock {\em Cognitive Processing}, 20(2):243--259.

\bibitem[Körber, 2019]{bagnara_theoretical_2019}
Körber, M. (2019).
\newblock Theoretical {Considerations} and {Development} of a {Questionnaire} to {Measure} {Trust} in {Automation}.
\newblock In Bagnara, S., Tartaglia, R., Albolino, S., Alexander, T., and Fujita, Y., editors, {\em Proceedings of the 20th {Congress} of the {International} {Ergonomics} {Association} ({IEA} 2018)}, volume 823, pages 13--30. Springer International Publishing, Cham.

\bibitem[Lee and Moray, 1992]{lee_trust_1992}
Lee, J. and Moray, N. (1992).
\newblock Trust, control strategies and allocation of function in human-machine systems.
\newblock {\em Ergonomics}, 35(10):1243--1270.

\bibitem[Lee and See, 2004]{lee_trust_2004}
Lee, J.~D. and See, K.~A. (2004).
\newblock Trust in {Automation}: {Designing} for {Appropriate} {Reliance}.
\newblock {\em Human Factors: The Journal of the Human Factors and Ergonomics Society}, 46(1):50--80.

\bibitem[Liu et~al., 2021]{liu_clustering_2021}
Liu, J., Akash, K., Misu, T., and Wu, X. (2021).
\newblock Clustering {Human} {Trust} {Dynamics} for {Customized} {Real}-{Time} {Prediction}.
\newblock In {\em 2021 {IEEE} {International} {Intelligent} {Transportation} {Systems} {Conference} ({ITSC})}, pages 1705--1712, Indianapolis, IN, USA. IEEE.

\bibitem[Mayer et~al., 1995]{mayer_integrative_1995}
Mayer, R.~C., Davis, J.~H., and Schoorman, F.~D. (1995).
\newblock An {Integrative} {Model} of {Organizational} {Trust}.
\newblock {\em The Academy of Management Review}, 20(3):709.

\bibitem[Parasuraman and Riley, 1997]{parasuraman_humans_1997}
Parasuraman, R. and Riley, V. (1997).
\newblock Humans and {Automation}: {Use}, {Misuse}, {Disuse}, {Abuse}.
\newblock {\em Human Factors: The Journal of the Human Factors and Ergonomics Society}, 39(2):230--253.

\bibitem[Parasuraman et~al., 2000]{parasuraman_model_2000}
Parasuraman, R., Sheridan, T., and Wickens, C. (2000).
\newblock A model for types and levels of human interaction with automation.
\newblock {\em IEEE Transactions on Systems, Man, and Cybernetics - Part A: Systems and Humans}, 30(3):286--297.

\bibitem[Patil et~al., 2010]{patil_pymc_2010}
Patil, A., Huard, D., and Fonnesbeck, C.~J. (2010).
\newblock {PyMC}: {Bayesian} {Stochastic} {Modelling} in {Python}.
\newblock {\em Journal of Statistical Software}, 35(4):1--81.

\bibitem[Ratcliff and Childers, 2015]{ratcliff_individual_2015}
Ratcliff, R. and Childers, R. (2015).
\newblock Individual differences and fitting methods for the two-choice diffusion model of decision making.
\newblock {\em Decision}, 2(4):237--279.

\bibitem[Razin and Feigh, 2023]{razin_converging_2023}
Razin, Y.~S. and Feigh, K.~M. (2023).
\newblock Converging {Measures} and an {Emergent} {Model}: {A} {Meta}-{Analysis} of {Human}-{Automation} {Trust} {Questionnaires}.

\bibitem[Rodriguez~Rodriguez et~al., 2023]{rodriguez_rodriguez_review_2023}
Rodriguez~Rodriguez, L., Bustamante~Orellana, C.~E., Chiou, E.~K., Huang, L., Cooke, N., and Kang, Y. (2023).
\newblock A review of mathematical models of human trust in automation.
\newblock {\em Frontiers in Neuroergonomics}, 4:1171403.

\bibitem[Rodríguez, 2010]{rodriguez_survival_2010}
Rodríguez, G. (2010).
\newblock Survival {Models}.

\bibitem[Smith, 2021]{smith_isaac_2021}
Smith, T. (2021).
\newblock {ISAAC}: {An} {Integrated} {System} for {Autonomous} and {Adaptive} {Caretaking}.

\bibitem[Tourangeau and Rasinski, 1988]{tourangeau_cognitive_1988}
Tourangeau, R. and Rasinski, K.~A. (1988).
\newblock Cognitive processes underlying context effects in attitude measurement.
\newblock {\em Psychological Bulletin}, 103(3):299--314.

\bibitem[Williams et~al., 2021]{williams_deconstructed_2021}
Williams, T., Ayers, D., Kaufman, C., Serrano, J., and Roy, S. (2021).
\newblock Deconstructed {Trustee} {Theory}: {Disentangling} {Trust} in {Body} and {Identity} in {Multi}-{Robot} {Distributed} {Systems}.
\newblock In {\em Proceedings of the 2021 {ACM}/{IEEE} {International} {Conference} on {Human}-{Robot} {Interaction}}, pages 262--271, Boulder CO USA. ACM.

\bibitem[Xu and Dudek, 2015]{xu_optimo_2015}
Xu, A. and Dudek, G. (2015).
\newblock {OPTIMo}: {Online} {Probabilistic} {Trust} {Inference} {Model} for {Asymmetric} {Human}-{Robot} {Collaborations}.
\newblock In {\em Proceedings of the {Tenth} {Annual} {ACM}/{IEEE} {International} {Conference} on {Human}-{Robot} {Interaction}}, pages 221--228, Portland Oregon USA. ACM.

\end{thebibliography}

\begin{biogs}

Jason Dekarske
\begin{itemize}
    \item Current Affiliation: University of California, Davis
    \item Highest Degree Obtained: PhD Mechanical and Aerospace Engineering, University of California, Davis, 2023
\end{itemize}

Gregory Bales
\begin{itemize}
    \item Current Affiliation: University of California, Davis
    \item Highest Degree Obtained:  PhD Mechanical and Aerospace Engineering, University of California, Davis, 2023
\end{itemize}

Zhaodan Kong
\begin{itemize}
    \item Current Affiliation: University of California, Davis
    \item Highest Degree Obtained:  PhD Aerospace Engineering and Mechanics, University of Minnesota, Twin Cities, 2012
\end{itemize}

Sanjay Joshi
\begin{itemize}
    \item Current Affiliation: University of California, Davis
    \item Highest Degree Obtained: PhD Electrical Engineering, University of California, Los Angeles, 1996
\end{itemize}
\end{biogs}


\end{document}